\documentclass[a4paper,aps,prd,preprint,longbibliography,nofootinbib,superscriptaddress]{revtex4-1}

\usepackage[T1]{fontenc}
\usepackage[utf8]{inputenc}
\usepackage{array}
\usepackage{amsmath,amssymb,amsfonts}
\usepackage{multirow}
\usepackage{float}
\usepackage{graphicx}
\usepackage{subfigure}

\usepackage[colorlinks=true, pdfstartview=FitV, linkcolor=blue, citecolor=blue, urlcolor=blue]{hyperref}
\allowdisplaybreaks[4]

\newcommand{\bw}{\begin{widetext}}
\newcommand{\ew}{\end{widetext}}
\newcommand{\be}{\begin{equation}}
\newcommand{\en}{\end{equation}}
\newcommand{\bee}{\begin{equation}}
\newcommand{\ene}{\end{equation}}
\newcommand{\bea}{\begin{eqnarray}}
\newcommand{\ena}{\end{eqnarray}}
\newcommand{\bes}{\begin{subequations}}
\newcommand{\ens}{\end{subequations}}
\newcommand{\bef}{\begin{figure}}
\newcommand{\enf}{\end{figure}}

\def\fb{\, {\rm fb}}

\def\ab{\, {\rm ab}}

\def\ie{{\it i.e.}}

\def\cala{\mathcal{A}}
\def\calb{\mathcal{B}}

\def\calm{\mathcal{M}}

\def\calo{\mathcal{O}}

\def\to{\rightarrow}

\def\gev{{\rm GeV}}

\begin{document}


\title{Testing Bell inequality through $h\to\tau\tau$ at CEPC}

\author{Kai Ma}
\email[]{makai@ucas.ac.cn}
\affiliation{Faculty of Science, Xi'an University of Architecture and Technology, Xi'an, 710055, China}

\author{Tong Li~\footnote[1]{corresponding author}}
\email[]{litong@nankai.edu.cn}
\affiliation{School of Physics, Nankai University, Tianjin 300071, China}

\date{\today}

\begin{abstract}
The decay of Higgs boson into two spin-1/2 particles provides an ideal system to reveal quantum entanglement and Bell-nonlocality. Future $e^+e^-$ colliders can improve the measurement accuracy of the spin correlation of tau lepton pairs from Higgs boson decay.
We show the testability of Bell inequality through $h\to \tau\tau$ at Circular Electron Positron Collider (CEPC). Two realistic methods of testing Bell inequality are investigated, i.e., T\"{o}rnqvist's method and Clauser-Home-Shimony-Holt (CHSH) inequality. In the simulation, we take into account the detector effects of CEPC including uncertainties for tracks and jets from $Z$ boson in the production of $e^+e^-\to Zh$. Necessary reconstruction approaches are described to measure quantum entanglement between $\tau^+$ and $\tau^-$. Finally, we show the sensitivity of CEPC to the Bell inequality violation for the two methods.
\end{abstract}

\maketitle

\tableofcontents

\section{Introduction}
\label{sec:intro}

It is well known that the most important debate on whether the Quantum Mechanics (QM) is a complete local theory is the challenge raised by Einstein, Podolsky, and Rosen (EPR), named EPR paradox~\cite{Einstein:1935rr}. The interpretation of the EPR paradox in local
hidden variable theory (LHVT) shows the contradiction of
LHVT with QM and presents the non-local nature of QM. Later on, Bohm et al. proposed a realistic experiment with a system of two spin-1/2 particles to illustrate the EPR paradox~\cite{Bohm:1957zz}. Based on this consideration, Bell established a theorem that the
two particles' spin correlation satisfies a Bell inequality (BI) in realistic LHVT~\cite{Bell:1964kc}. By contrast, the QM predictions may violate this inequality in some certain parameter space. Clauser, Horne, Shimony and Holt (CHSH) also generalized the original Bell inequality and established a more practical inequality~\cite{Clauser:1969ny}. The test of the Bell inequality delivers a direct justification if the QM is a complete local theory~\cite{Bertlmann:2004yg}.

In the past years, the violation of this Bell inequality has been observed in many low-energy experiments (such as optical experiments)~\cite{Aspect:1981nv,Aspect:1982fx,Ou:1988zz,Weihs:1998gy,Bouwmeester:1998iz,Pan1,PhysRevLett.83.3103,PhysRevLett.95.240406} as the foundation of the modern quantum information theory. The predictions of QM are proved to be consistent with the results of these experiments.
However, for testing the completeness of QM beyond the electromagnetic interaction regime, it is still a challenge of the test of Bell inequality in high-energy physics (see review Ref.~\cite{Ding:2007mk} and references therein). At $e^+e^-$ colliders, the testability of BI was first suggested by using the polarization correlation in the process of $e^+e^-\to \Lambda \bar{\Lambda}\to \pi^- p \pi^+ \bar{p}$~\cite{Tornqvist:1980af,Hao:2009kj} or $e^+e^-\to Z\to \tau^+\tau^-$~\cite{Privitera:1991nz,Abel:1992kz,Dreiner:1992gt}.
Based on the CHSH method, dedicated proposals were also raised to test the BI in the final states of $t\bar{t}$ pair~\cite{Afik:2020onf,Fabbrichesi:2021npl,Severi:2021cnj,Afik:2022kwm,Aguilar-Saavedra:2022uye,Afik:2022dgh,Aoude:2022imd,Fabbrichesi:2022ovb,Varma:2023gwh,Dong:2023xiw} or two weak gauge bosons~\cite{Barr:2021zcp,Barr:2022wyq,Aguilar-Saavedra:2022mpg,Aguilar-Saavedra:2022wam,Fabbrichesi:2023cev,Aoude:2023hxv,Bernal:2023ruk,Fabbri:2023ncz,Bi:2023uop} at the Large Hadron Collider (LHC).
Nevertheless, the spin-0 state formed by a pair of spin-1/2 particles in the process such as $e^+e^-\to \Lambda \bar{\Lambda}\to \pi^- p \pi^+ \bar{p}$ has the largest entanglement.
It is because the two spin-1/2 particles emerging from the decay of a spin-0
particle always have the same helicity. In contrast, for the production in a mixture of
spin singlet and triplet channels, the magnitude of the correlation
is smaller because of the cancellation of different contributions.

The Higgs boson is the only spin-0 elementary particle in the Standard Model (SM) and can play as a natural spin singlet state to test LHVT through the Bell inequality at high energies. The properties of SM Higgs boson will be measured to high precision at future $e^+e^-$ colliders such as the Circular Electron Positron Collider (CEPC)~\cite{CEPCStudyGroup:2018ghi}.
We thus propose to test the Bell inequality at CEPC through the Higgsstrahlung process with subsequent decay $h\to \tau^+\tau^-$~\cite{Altakach:2022ywa,Fabbrichesi:2022ovb}
\begin{eqnarray}
e^+e^-\to Zh\to Z\tau^+\tau^-\;.
\end{eqnarray}
The tau lepton pair is correlated in the decay process, and Bell inequality
(or quantum entanglement) can be tested by measuring their spin correlation.
However, spin information of the tau leptons can only be partially inferred from its decay particle.
Here we consider only the tau leptons followed by the 1-prong decay mode
$\tau^\pm \to \pi^\pm \nu_\tau$ which is the best spin analyzer for the tau lepton polarization.
In principle, the other decay modes can also be employed. However, it is more challenging in practice
because of the kinematic reconstruction of the tau lepton as well as limited spin analyzing power.
In order to have higher statistics,
the associated $Z$ boson will be reconstructed by both its leptonic and hadronic decay modes.
Furthermore, both T\"{o}rnqvist's method~\cite{Tornqvist:1980af} and the CHSH method~\cite{Clauser:1969ny} are explored to evaluate the violation of Bell inequality.
To have a more realistic estimation on the experimental sensitivities, we investigate the kinematic reconstruction and simulate the detector effects to reveal the quantum entangled spin correlations in the decay of tau pairs.

Note that QM predicts a larger bound of the joint expectation values in CHSH method, that is $2\sqrt{2}$ but not 2~\cite{Cirelson:1980ry}. One may suspect a certain QM setup would not violate the Bell inequality. New physics beyond the SM may modify the angular distributions
of the tau-lepton decay products. However, such kind of new physics has no solid evidence up to now and is restricted to be very small
compared to the SM interactions.
Therefore, our work is conducted based on
the assumption that both the $h\to \tau\tau$ decay and tau decay processes
are described by the SM.

This paper is organized as follows. In Sec.~\ref{sec:Bell}, we first outline the LHVT and Bell inequality.
Then we show T\"{o}rnqvist's method and the CHSH method in terms of the polarization correlation in decay $h\to \tau^+\tau^-\to \pi^+\bar{\nu}_\tau \pi^- \nu_\tau$. In Sec.~\ref{sec:LC}, we describe the simulation of process $e^+e^-\to Zh\to Z\tau^+\tau^-$ and discuss the detector effects as well as reconstruction methods.
The results of projected sensitivity to the Bell inequality violation are given in Sec.~\ref{sec:Sen}.
Finally, in Sec.~\ref{sec:Con} we summarize our conclusions.

\section{Local Quantum Model and Bell Inequality}
\label{sec:Bell}

In this section, we describe the original and generalized expressions of Bell inequality and the realistic methods of testing it in high-energy physics.

In the LHVT with the hidden variable being $\lambda$, the Bell inequality can be phased in terms of the polarization correlation
\bee
P(\vec{a}, \vec{b})
=
\int d\lambda \, q(\lambda) \cdot
\mathcal{P}_{A}(\vec{a}, \lambda)\cdot \mathcal{P}_{B}(\vec{b}, \lambda)\,,
\ene
where $\mathcal{P}_{A(B)}(\vec{x}, \lambda)$ is the probability of the fermion $A$ (or $B$) with
spin along the direction $\vec{x}=\vec{a}~(\vec{b})$ for given hidden variable $\lambda$, and
$q(\lambda)$ is the corresponding probability distribution of the
hidden variable $\lambda$.
The original expression of Bell inequality refers to three independent
spatial directions $\vec{a}$, $\vec{b}$ and $\vec{c}$ as
\bee
\label{eq:bell:3v}
\big| P(\vec{a}, \vec{b}) - P(\vec{a}, \vec{c})  \big|
\le 1 + P(\vec{b}, \vec{c})\;.
\ene
On the other hand, in QM the quantum average of the correlation operator $\calo(\vec{a}, \vec{b})
\equiv
\big[ \vec{\sigma}^A \cdot \vec{a} \big] \;
\big[ \vec{\sigma}^B \cdot \vec{b} \,\big]$ is given by
\bee
P(\vec{a}, \vec{b})
=
\big\langle 00 \big| \big[ \vec{\sigma}^A \cdot \vec{a} \big] \;
\big[ \vec{\sigma}^B \cdot \vec{b} \,\big] \,\big| 00 \big\rangle
=
- \vec{a} \cdot \vec{b}\,,
\label{eq:QM}
\ene
where $\langle 00|$ or $|00\rangle$ refers to a singlet state of the total spin.
After inserting the QM prediction Eq.~(\ref{eq:QM}) in Eq.~(\ref{eq:bell:3v}), the Bell inequality Eq.~(\ref{eq:bell:3v}) may be violated in some region of phase space. However, in realistic investigations, the spin correlation of the two fermions $A$ and $B$ can only be transferred to the kinematics of their decay products. In terms of $h\to \tau^+\tau^-$ and tau leptons' hadronic decay mode $\tau^\pm\to \pi^\pm \nu_\tau$, we will describe two existing methods to perform the test of Bell inequality at high energy colliders.

\subsection{T\"{o}rnqvist's method}

In Ref.~\cite{Tornqvist:1980af}, T\"{o}rnqvist suggested to test the BI by using the polarization correlation in the process
\begin{eqnarray}
e^+e^-\to \Lambda \bar{\Lambda}\to \pi^- p \pi^+ \bar{p}\;.
\end{eqnarray}
The parent particle of $\Lambda \bar{\Lambda}$ could be either spin-0 $\eta_c$ or spin-1 $J/\psi$.
We instead establish the polarization correlation of decay $h\to \tau^+\tau^-\to \pi^+\bar{\nu}_\tau \pi^- \nu_\tau$.
Since the Higgs boson is a scalar and the $\tau$-lepton is a spin-1/2 particle,
the decay process $h\to \tau^+\tau^-$ provides an ideal system for testing
the Bell inequality.
The joint spin density matrix for the $\tau^+\tau^-$ system is given by
\bee
\rho_{\tau\bar\tau}
=
\frac{1}{4} \Big( 1 - \vec{\sigma}_{\tau} \cdot \vec{\sigma}_{\bar\tau} \Big)\,,
\ene
which means the state with parallel $\vec{\sigma}_{\tau}$ and $\vec{\sigma}_{\bar\tau}$ vanishes because of spin-zero condition.
For the correlation operator $\calo(\vec{a}, \vec{b})$,
one can easily find the probability is given as
\bee
P(\vec{a}, \vec{b})
=
\big\langle 00 \big| \rho_{\tau\bar\tau}  \calo(\vec{a}, \vec{b}) \big| 00 \big\rangle
=
- \vec{a} \cdot \vec{b}\,.
\ene
We show the calculation of the spin correlation coefficients in Appendix~\ref{Appendix}.

However, the spin states of the $\tau$-leptons can not be measured directly at collider,
and can only be accessed by the angular distributions of their decay products.
Here we only investigate the 1-prong decay mode $\tau^- \to \pi^- \nu_\tau$,
in which the momentum direction of the charged pion (or equivlently the neutrino) is
correlated to the spin direction of the tau lepton. Thus, this decay mode has the largest spin analyzing power compared to the cases of the other decay modes.
The decay amplitude of the process $\tau^- \to \pi^- \nu_\tau$ in the rest frame
of the mother particle can be written as
\bee
\calm_{\tau}
=
\frac{1}{\sqrt{ 4\pi }} \big( S + P \vec{\sigma}_\tau \cdot \vec{a} \big)\,,
\ene
where $\vec{a}$ is the unit vector along the $\pi^-$ momentum direction
in the rest frame of $\tau^-$,
$S$ and $P$ are the $S$- and $P$-wave amplitudes respectively.
Similar expression is valid for decay process $\tau^+ \to \pi^+ \bar{\nu}_\tau$ as well.
Then, the probability of having $\pi^-$ flying along $\vec{a}$ and
$\pi^+$ flying along $\vec{b}$ ($\vec{b}$ is the unit vector along the $\pi^+$ momentum direction
in the rest frame of $\tau^+$) becomes
\bee
\widetilde{P}(\vec{a}, \vec{b})
=
\big\langle 00 \big| \rho_{\tau\bar\tau}  \big[ \calm_{\tau} \calm_{\bar\tau}  \big]^\dag \;
\big[ \calm_{\tau} \calm_{\bar\tau}  \,\big] \,\big| 00 \big\rangle
=
\left[ \frac{1}{4\pi} \Big( \big| S \big|^2 +  \big| P \big|^2 \Big)  \right]^2
\Big( 1 + \alpha^2 \vec{a} \cdot \vec{b} \Big)\,,
\ene
where
\bee
\alpha = - \frac{ 2\Re S P^\ast }{ \big| S \big|^2 +  \big| P \big|^2 } \approx 0.573\,.
\ene
The above value is obtained by fitting in our numerical simulation.
One can see that $\widetilde{P}(\vec{a}, \vec{b})$ is a partial measurement of
the spin states of the $\tau$-lepton pair. Its normalized value $\widetilde{P}^N(\vec{a}, \vec{b})$ is related to $P(\vec{a}, \vec{b})$ by the following relation
\bee
P(\vec{a}, \vec{b})
=
\frac{1}{ \alpha^2 } \Big[ 1 - \widetilde{P}^N(\vec{a}, \vec{b}) \Big]\,.
\ene
The normalized differential cross section is given as
\bee
\frac{1}{\sigma} \frac{ d\sigma}{d\cos\theta_{ab}}
= {1\over 2}\widetilde{P}^N(\vec{a}, \vec{b})
= \frac{1}{2} \left[ 1 - \alpha^2 \, P( \vec{a}, \vec{b} ) \right] \,,
\label{eq:TtestQM}
\ene
where $\cos\theta_{ab} \equiv \vec{a} \cdot \vec{b}=-P( \vec{a}, \vec{b} )$.
On the other hand, hidden variable theory predicts~\cite{Tornqvist:1980af}
\bee
\left| P( \vec{a}, \vec{b} ) \right| \le 1 - \frac{2}{\pi} \theta_{ab}\,,~~\theta_{ab} \in [0, \pi] \;.
\label{eq:LHVT}
\ene
Then we have the following classical region satisfying the Bell inequality
\bee
\begin{cases}
\dfrac{1}{2} - \alpha^2 \Big(\dfrac{1}{2}- \dfrac{\theta_{ab}}{\pi}\Big)
\le\dfrac{1}{\sigma} \dfrac{ d\sigma}{d\cos\theta_{ab}}
\le
\dfrac{1}{2} + \alpha^2 \Big(\dfrac{1}{2}- \dfrac{\theta_{ab}}{\pi}\Big)\;,  & \theta_{ab} \in [0, \pi/2]
\\[3mm]
\dfrac{1}{2} + \alpha^2 \Big(\dfrac{1}{2}- \dfrac{\theta_{ab}}{\pi}\Big)
\le\dfrac{1}{\sigma} \dfrac{ d\sigma}{d\cos\theta_{ab}}
\le
\dfrac{1}{2} - \alpha^2 \Big(\dfrac{1}{2}- \dfrac{\theta_{ab}}{\pi}\Big)\;,  & \theta_{ab} \in (\pi/2, \pi]
\end{cases}
 \,.
 \label{eq:TtestLHVT}
\ene

\subsection{Clauser-Home-Shimony-Holt Inequality}

Clauser, Horne, Shimony and Holt (CHSH) generalized the original Bell inequality
Eq.~(\ref{eq:bell:3v}) by considering general properties of the quantum density
matrix of a spin-1/2 particles system~\cite{Clauser:1969ny}.
Density matrix of the quantum state having two spin-1/2 particles
can be expressed in general as
\bee
\rho
=
\frac{1}{4} \left[ \mathbb{I}_A \otimes \mathbb{I}_B
+ A_i \cdot \big( \sigma_{A,i} \otimes  \mathbb{I}_B \big)
+ B_j \cdot \big( \mathbb{I}_A \otimes  \sigma_{B,j} \big)
+
C_{ij} \big(  \sigma_{A,i} \otimes  \sigma_{B,j} \big)
\right]
\ene
where $\sigma_{A(B), i}$ and $\mathbb{I}_{A(B)}$ are Pauli matrices
and the unit $2\times2$ matrix for the particle $A~(B)$, respectively.
The Bell operator associated with the quantum CHSH inequality is
defined as
\bee
\calb_{\rm CHSH}
=
\vec{a} \cdot \vec{\sigma}_A \otimes \big( \vec{b} + \vec{b}' \big)  \cdot \vec{\sigma}_B
+
\vec{a}' \cdot \vec{\sigma}_A \otimes \big( \vec{b} - \vec{b}' \big)  \cdot \vec{\sigma}_B \,,
\ene
where $\vec{a}$, $\vec{a}'$, $\vec{b}$, $\vec{b}'$ are unit vectors.
Then the CHSH inequality is given by~\cite{HORODECKI1995340}
\bee
\Big| {\rm Tr}(\rho \calb_{\rm CHSH} ) \Big| \le 2\;.
\ene
Again, in practice it is hard to test the above inequality directly because of
the challenge in measuring of the spin directions $\vec{a}$, $\vec{a}'$, $\vec{b}$, $\vec{b}'$. Alternatively,
the matrix with following coefficients
\bee
C_{ij} = {\rm Tr} \big[ \rho \sigma_i \otimes \sigma_j \big] \,,
\ene
can provide an indirect inequality.
It was shown that if sum of the two largest eigenvalues of the matrix
$U = C^T C$ is larger than 1, then the CHSH inequality is violated~\cite{HORODECKI1995340}.

At colliders, the density matrix can be estimated
by angular distributions of the two spin-1/2 particles' decay products.
The normalized differential cross section can be generally parameterized
as~\cite{Bernreuther:2015yna}
\bee
\frac{ \sigma^{-1} d\sigma}{ d\cos\theta_{A,i } d\cos\theta_{B,j } }
=
\frac{1}{4} \Big[ 1
+ A_i \cos\theta_{A,i } + B_j \cos\theta_{B,j } + C_{ij} \cos\theta_{A,i } \cos\theta_{B,j }
\Big]\;,
\label{eq:xsec}
\ene
where $\theta_{A(B),i(j)}$ are the polar angle of charged particles $A~(B)$ from the decays
of their mother particles, and measured from the $i(j)$-th axis. The helicity basis is always chosen for the spins of the two
taus.
In our case of $h\to \tau^+\tau^-\to \pi^+\bar{\nu}_\tau \pi^- \nu_\tau$, the cosine quantities of the above polar angles are defined as
\begin{eqnarray}
\cos\theta_{\pi^+,i}=\hat{p}_{\pi^+}\cdot \hat{i}\;,~~~\cos\theta_{\pi^-,j}=\hat{p}_{\pi^-}\cdot \hat{j}\;,
\end{eqnarray}
where the unit vectors $\hat{i}$ and $\hat{j}$ are defined in the rest frames of the
$\tau^+$ and $\tau^-$, respectively. They belong to a chosen orthonormal basis $\hat{j}\in \{\hat{k},\hat{r},\hat{n}\}$ and satisfy the relation $\hat{i}=-\hat{j}$. More precisely, we define an unit vector $\hat{k}$ as the direction of $\tau^-$ momentum in the rest frame of the Higgs boson. In the rest frame of the $\tau^-$ lepton, we define
an unit vector $\hat{r}$ in the decay plane of the $\tau^-$ lepton and perpendicular to $\hat{k}$,
and an unit vector $\hat{n} = \hat{k}\times \hat{r}$.
It was shown that the matrix $C$ can be calculated as~\cite{Bernreuther:2015yna,Fabbrichesi:2021npl}.
\bee
C_{ij}
=
- 9 \int  d\cos\theta_{A,i } d\cos\theta_{B,j } \frac{ \sigma^{-1} d\sigma}{ d\cos\theta_{A,i } d\cos\theta_{B,j } }
\cos\theta_{A,i } \cos\theta_{B,j } \;.
\ene
Then, one can diagonalize the spin correlation matrix $C^TC$ and find the two largest eigenvalues to test the CHSH inequality.
Since the Higgs boson is considered to be on-shell, and its spin is zero,
there is no any invariant mass and orientation dependencies, compared to the $t\bar{t}$ final states in
Ref.~\cite{Fabbrichesi:2021npl}.

\section{Measurements at Future Lepton Colliders}
\label{sec:LC}

At $e^+e^-$ colliders, the dominant production mode of the Higgs boson is the so-called
Higgsstrahlung channel, $e^+e^- \to Z h$. For our interested mode $h\to\tau^+\tau^-$
with subsequent decay channels $\tau^\pm \to \pi^\pm \nu$, two neutrinos appear in the final
state. Hence kinematic reconstruction is necessary in order to measure quantum
entanglement between $\tau^+$ and $\tau^-$.
Next, we describe our numerical simulation and implementation of the detector effects,
and then the reconstruction approaches in both the leptonic and hadronic decay modes of $Z$ boson.

\subsection{Simulation and Detector Effects}

Our numerical simulations are conducted using the
\texttt{MadGraph5\_aMC@NLO}~\cite{Alwall:2014hca} package, and the quantum entangled
spin correlations in the tau-lepton decay are preserved by
the TauDecay~\cite{Hagiwara:2012vz} package.
For a realistic simulation, detector resolutions have to be included for the objects.
Charged tracks can be precisely measured by the CEPC detector for the decay products in $Z\to \ell^+\ell^-~(\ell=e,\mu)$ or $\tau^\pm\to \pi^\pm \nu$.
Table~\ref{table:sigmaCEPC} lists typical uncertainties of the azimuthal angle ($\phi$),
rapidity ($\eta$) and magnitude of the transverse momentum ($|\vec{p}_T|$) at CEPC~\cite{CEPCStudyGroup:2018ghi}.
One can see that the CEPC uncertainties for tracks are quite small.
We smear tracks (both leptons and pions) by randomly sampling the azimuthal angle,
pseudo-rapidity and transverse momentum according to Gaussian distribution
with standard deviations given in Table~\ref{table:sigmaCEPC}~\cite{deFavereau:2013fsa,LinearColliderILDConceptGroup-:2010nqx}.

\begin{table}[thb]
\renewcommand\arraystretch{1.22}
\begin{center}
\begin{tabular}{cc}
\hline
\makebox[0.2\textwidth]{Observables}  & \makebox[0.28\textwidth]{Uncertainties}
\\ \hline
$\phi$ & 0.0002$|\eta|$ + 0.000022
\\ \hline
$\eta$ & 0.000016$|\eta|$ + 0.00000022
\\ \hline
$|\vec{p}_T|$ & 0.036$|\vec{p}_T|$
\\
\hline
\end{tabular}
\end{center}
\caption{CEPC uncertainties for tracks}
\label{table:sigmaCEPC}
\end{table}

However, uncertainties of jet from $Z$ boson's hadronic decay are relatively large.
Measurement of jet is not only smeared by fragmentation of partons and the corresponding
jet clustering processes, but also by the jet clustering of reconstructed objects after detector
response and its matching to jet at the generator level~\cite{Ruan:2018yrh}.
The results in Ref.~\cite{Lai:2017rko,Lai:2021rko} indicate that
the uncertainty induced by the jet clustering and matching can be as
significant as those from the detector response, and becomes the dominant
uncertainty especially for final state with more than two jets.
Hence, sophisticated jet clustering algorithm has to be used~\cite{Ruan:2018yrh}.
The energy resolution of the jet from light quarks can be described as~\cite{CEPCStudyGroup:2018ghi}
\bee
\label{eq:sigmaJet}
\sigma_{\rm jet}(E) = \frac{25.7\%}{\sqrt{E}} \, \oplus \, 2.4\% \,.
\ene
Jets from charm and bottom quarks have slightly larger uncertainties
because of neutrinos in their decays~\cite{CEPCStudyGroup:2018ghi}.
In consideration of this, in this paper we also use a smearing algorithm to account
for detector resolutions.
Detector responses to the partons (for the channel $Z\to q\bar{q}$) are included
by smearing energy of the partons according to Gaussian
distribution with standard deviations given in Eq.~\eqref{eq:sigmaJet}.

\begin{figure}[htb!]
\centering
\includegraphics[width=0.32\textwidth]{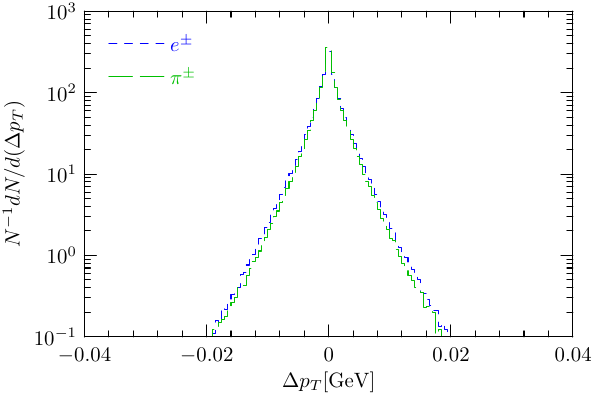}
\includegraphics[width=0.32\textwidth]{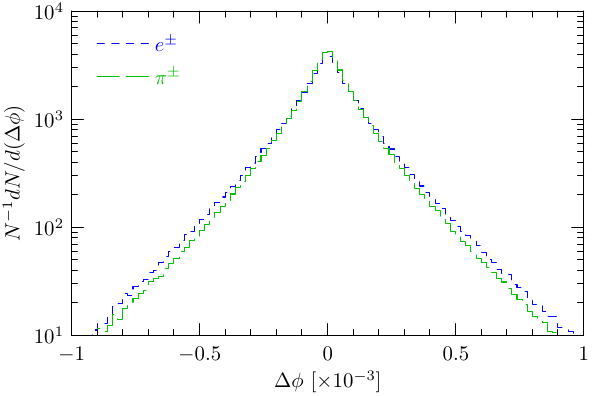}
\includegraphics[width=0.32\textwidth]{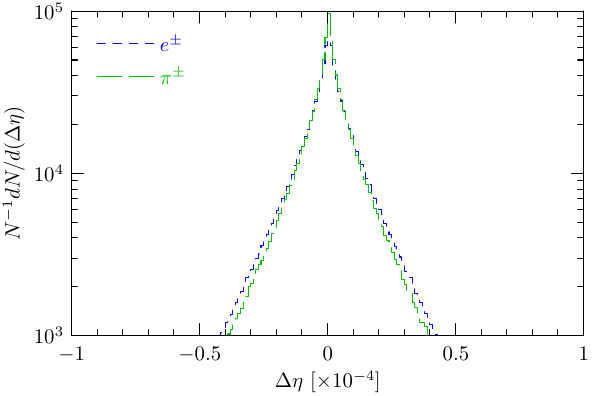}\\
\includegraphics[width=0.32\textwidth]{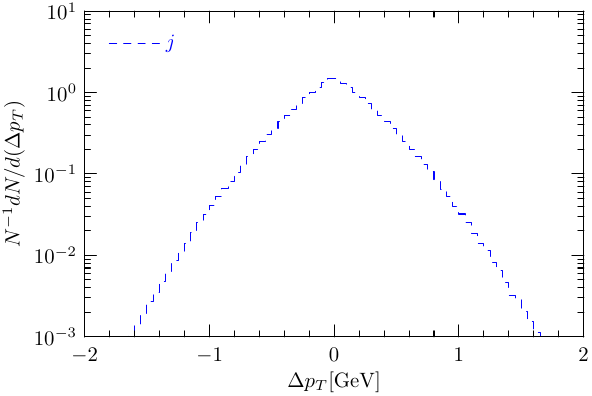}
\includegraphics[width=0.32\textwidth]{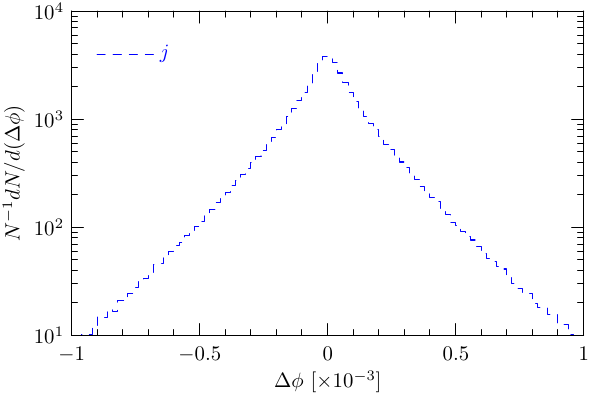}
\includegraphics[width=0.32\textwidth]{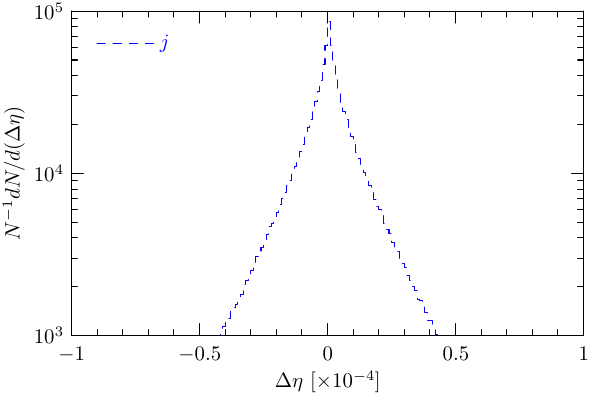}
\caption{
Normalized number of events as a function of $\Delta p_T$ (left), $\Delta \phi$ (middle) and $\Delta \eta$ (right) for the objects (blue: $e^\pm$ or jet $j$, green: $\pi^\pm$) in leptonic (top) and hadronic (bottom) decay modes of $Z$ boson.
}
\label{fig:uncer}
\end{figure}

\begin{figure}[htb!]
\centering
\includegraphics[width=0.32\textwidth]{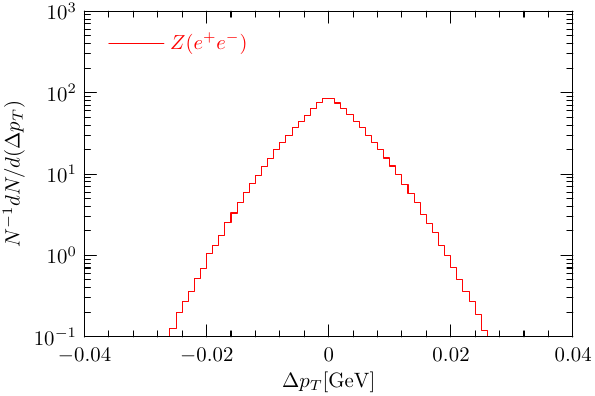}
\includegraphics[width=0.32\textwidth]{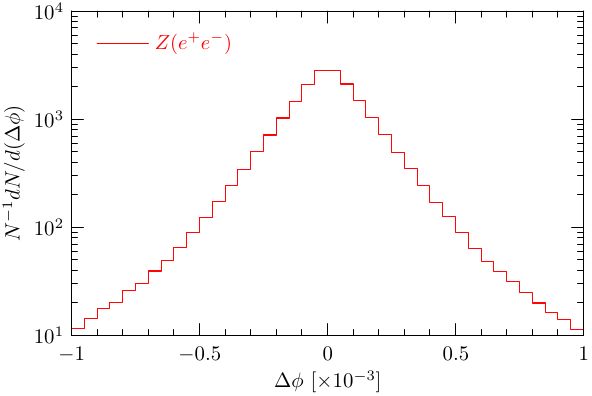}
\includegraphics[width=0.32\textwidth]{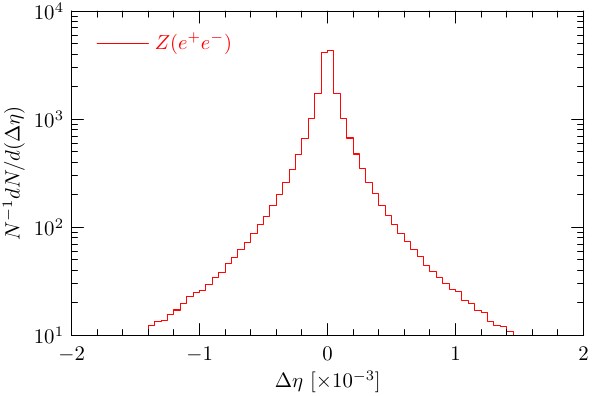}\\
\includegraphics[width=0.32\textwidth]{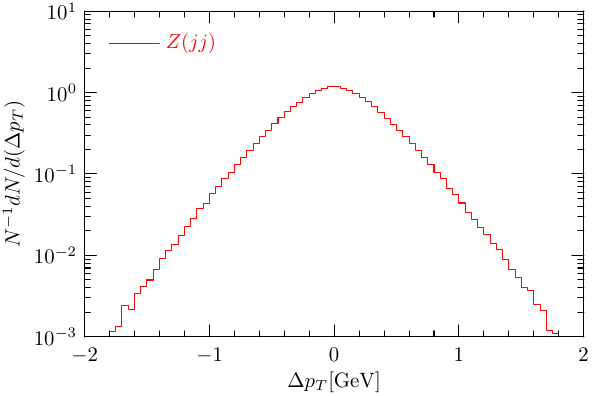}
\includegraphics[width=0.32\textwidth]{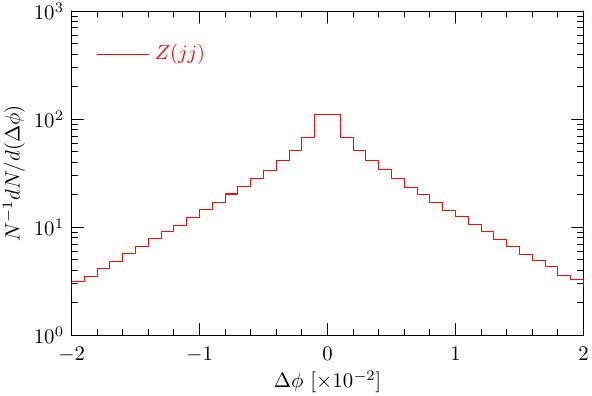}
\includegraphics[width=0.32\textwidth]{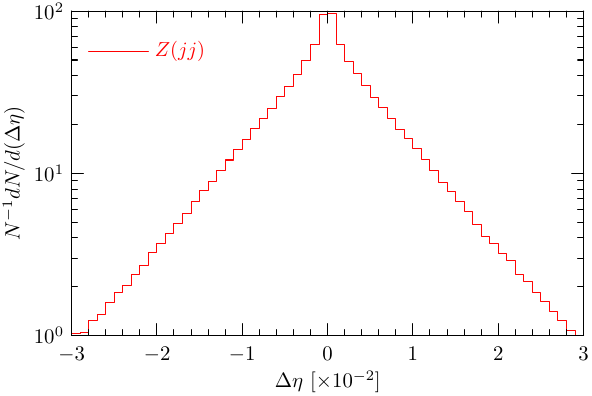}
\caption{
Normalized number of events as a function of $\Delta p_T$ (left), $\Delta \phi$ (middle) and $\Delta \eta$ (right) for the $Z$ boson in leptonic (top) and hadronic (bottom) decay modes.
}
\label{fig:uncerZ}
\end{figure}

To see the impact of the above uncertainties, in Fig.~\ref{fig:uncer}, we show the distributions of the difference between the real value and the smeared value of transverse momentum $p_T$, azimuthal angle $\phi$ and rapidity $\eta$ (defined as $\Delta p_T$, $\Delta \phi$ and $\Delta \eta$) for the objects in different decay modes of $Z$ boson. One can see that, due to the jet energy smearing, the $p_T$ uncertainties of jets in $Z$ boson's hadronic decay are quite large compared to those in the leptonic mode. As a result, the $Z$ boson decaying to dijet is not well reconstructed as shown in Fig.~\ref{fig:uncerZ}.

Furthermore, detection efficiency is also affected by particle identification.
For CEPC, the detector is designed to identify prompt
leptons with high efficiency and high purity~\cite{CEPCStudyGroup:2018ghi}.
For leptons with energies above 5~GeV, the identification efficiency is higher than 99\%
and misidentification rate is smaller than 2\%. For the $\tau$-jet with visible energy
between 20 and 80 GeV, the identification efficiency is above 80\% with a purity closing
to 90\%~\cite{CEPCStudyGroup:2018ghi}, further improvement can be expected by
optimizations. In our simulation, we ignore the momentum dependence and use an universal identification efficiency 80\% estimate experimental significance for $\tau$-jet.
For jets from hadronic decay of the $Z$ boson, b-jets can be tagged with an efficiency
of 80\% and a purity of 90\%. Similarly, an efficiency of 60\% and a purity of 60\% can
be achieved for the c-jet tagging~\cite{CEPCStudyGroup:2018ghi}. In our case,
since the $Z$ boson is treated inclusively, jet-tagging is irrelevant to our analysis.
Therefore, we will use an factor of $0.8$ to account for possible efficiency loss in
reconstruction at the detector level.

\subsection{Reconstruction Method}
\label{sec:recon}

For decay of the Higgs boson, the degree of
freedom of the corresponding phase space is 8, and 6 of them can be measured thanks to
the two charged pions. As a result, only 2 of them are undetermined. Considering that the
decay width of $\tau$-lepton is very small compared to its mass, it is an excellent
approximation to assume that both $\tau$s are on-shell. With help of the on-shell conditions,
the 8 kinematic degree of freedom can be determined. In the following studies, we always
assume that the $Z$-boson is reconstructed by its visible decay products, and momentum
of the Higgs boson is obtained by energy-momentum conservation condition,
\ie, $p_h = p_{e^+} + p_{e^-} - p_{Z}$. In the approximation of that $P^0 = \sqrt{s}$,
and $\vec{P} = 0$ with $P = p_{e^+} + p_{e^-} $, invariant mass of the Higgs momentum
is given by $p_h^2 = s + p_Z^2 - 2 E_{Z} \sqrt{s} $. Since the decay width of Higgs boson
is also expected to be very small, $p_h^2 \approx m_h^2$ is again an excellent approximation.
In practice, $p_h^2\equiv \hat{m}_h^2$ may deviate from $m_h^2$ significantly due to experimental
uncertainties in measurement of the $Z$ boson momentum.

Since the Higgs boson decay isotropically, the reconstruction is done in the rest frame of $h$.
Assuming that both $\tau$-leptons are on-shell,
then energy and magnitude of the $\tau$-lepton momentum in this reference frame
can be obtained directly,
\bee
E_{\tau}^\star = \frac{1}{2} \hat{m}_{h}\,,\;\;\;
p_{\tau}^\star = \frac{1}{2} \hat{m}_{h} \sqrt{ 1 - \frac{ 4 m_\tau^2 }{ \hat{m}_h^2 } } \,.
\ene
Intersection angles between momentum of the $\tau^{\pm}$ and $\pi^{\pm}$ in this frame
are given as,
\bee
\cos\theta_{\pm}^{\star}
=
\frac{ 2 E_{\tau}^\star E_{\pi^\pm}^\star - m_{\tau}^2 - m_{\pi^\pm}^2   }{ 2 p^\star_{\tau} p^\star_{\pi^\pm} }
\ene
Without loss of generality, we define the $z$-axis as the momentum direction of the negatively
charged decay product, and the positively charged decay product lies in the $x-z$ plane.
In this reference frame, momenta of the of the charged decay products can be written as,
\bea
p_{\pi^-}^{\star\mu}
&=&
E_{\pi^-}^{\star} \big(1,\, 0,\, 0,\, \beta_{\pi^-}^{\star} \big) \,,
\\[3mm]
p_{\pi^+}^{\star\mu}
&=&
E_{\pi^+}^{\star} \big(1,\,
\beta_{\pi^+}^{\star}\sin\theta_{\pi^+}^{\star},\,
0,\,
\beta_{\pi^+}^{\star}\cos\theta_{\pi^+}^{\star} \big)\,.
\ena
Furthermore, defining the azimuthal angle of the $\tau^-$-lepton as $\phi_{-}^{\star}$,
its momentum can be parameterized as
\bee
p_{\tau^-}^{\star\mu}
=
E_{\tau}^{\star} \big(1,\,
\beta_{\tau}^{\star}\sin\theta_{-}^{\star}\cos\phi_{-}^{\star},\,
\beta_{\tau}^{\star}\sin\theta_{-}^{\star}\sin\phi_{-}^{\star},\,
\beta_{\tau}^{\star}\cos\theta_{-}^{\star} \big) \,,
\ene
where $\beta_{\tau}^{\star} = \sqrt{1 - 4m_{\tau}^2/\hat{m}_h^2 }$.
It turns out that momentum of the $\tau^+$-lepton is given as
\bee
p_{\tau^+}^{\star\mu}
=
E_{\tau}^{\star} \big(1,\,
-\beta_{\tau}^{\star}\sin\theta_{-}^{\star}\cos\phi_{-}^{\star},\,
-\beta_{\tau}^{\star}\sin\theta_{-}^{\star}\sin\phi_{-}^{\star},\,
-\beta_{\tau}^{\star}\cos\theta_{-}^{\star} \big) \,.
\ene
Using the equation,
$
\vec{  p}_{\tau^+}^{\;\star} \cdot \vec{p}_{\pi^+}^{\;\star}
=
\cos\theta_{+}^{\star} \big|\vec{  p}_{\tau^+}^{\;\star} \big| \big| \vec{p}_{\pi^+}^{\;\star} \big|\,,
$
one can immediately have,
\bee
-\sin\theta_{\pi^+}^{\star}\sin\theta_{-}^{\star}
\cos\big(\phi_{-}^{\star} \big) 
-\cos\theta_{\pi^+}^{\star}\cos\theta_{-}^{\star}
=
\cos\theta_{+}^{\star} \,.
\ene
Then we get following solutions,
\bee
\phi_{-}^{\star} =
\pm
\arccos\left[ -
\frac{  \cos\theta_{\pi^+}^{\star}\cos\theta_{-}^{\star} + \cos\theta_{+}^{\star} }
{ \sin\theta_{\pi^+}^{\star}\sin\theta_{-}^{\star}   }
\right] \,.
\ene
Both solutions satisfy all the kinematic constraints, hence there is a two-fold ambiguity.

\begin{figure}[htb!]
\centering
\includegraphics[width=0.46\textwidth]{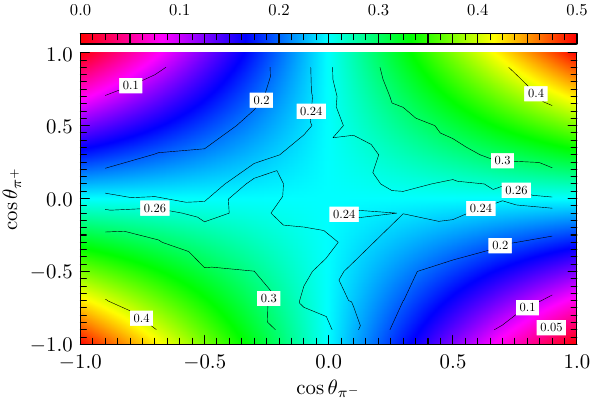}
\hfill%
\includegraphics[width=0.46\textwidth]{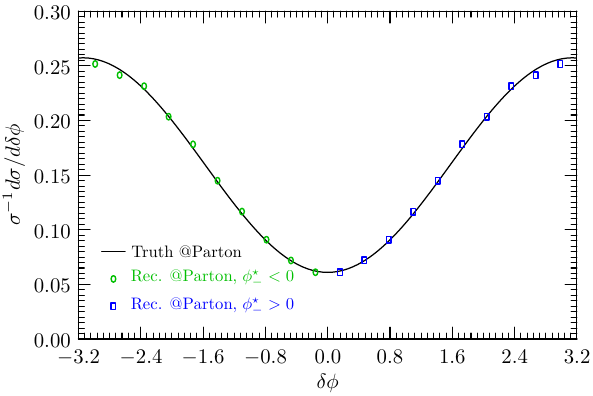}
\caption{
Longitudinal (left) and transverse (right) correlations as a result of analytical reconstruction method for leptonic decay mode of $Z$ boson at parton level.
}
\label{fig:Rec-True-parton}
\end{figure}

We then test the above analytical reconstruction method at parton level. Fig.~\ref{fig:Rec-True-parton} shows the longitudinal and transverse correlations as a result of the above analytical solutions for the leptonic decay mode of $Z$ boson. In the left panel of Fig.~\ref{fig:Rec-True-parton}, the colored densities indicate the true values of $\cos\theta_{\pi^+}$ versus $\cos\theta_{\pi^-}$ and the black contours show the reconstructed values.
The above two-fold ambiguity induces a reduction of the transverse
spin correction which is described by the azimuthal angle difference of the two decay planes
$\delta \phi$, as shown in the right panel of Fig.~\ref{fig:Rec-True-parton}.
In the leptonic decay mode of $Z$, one can see that the above analytical reconstruction method works well.

However, there are some drawbacks in this analytical reconstruction method. First of all, the two-fold ambiguity of kinematic solutions exists as mentioned above. One cannot determine the complete $\phi_{-}^{\star}$ distribution at a time. More importantly, due to the energy uncertainty of jets, the $Z$ boson cannot be well reconstructed in its hadronic decay mode. As a result, the uncertainties of the Higgs momentum given by
analytical reconstruction are very large in the hadronic decay mode of
the $Z$ boson. Quantum correlation effects are completely washed out,
and hence it is nearly impossible to observe violation of the Bell identity in
the hadronic mode. Therefore, we adopt the other reconstruction method by using impact parameters in next section.

\subsection{Reconstruction by using Impact Parameters}
\label{sec:impact}

It was shown that impact parameters
of the charged $\pi$s are very useful to reconstruct the full decay kinematics~\cite{Hagiwara:2016zqz,Jeans:2018anq,Chen:2017nxp}.
The $\tau$-leptons emerging from Higgs decay are strongly boosted.
As a result, its typical decay length $\sim 3000~\mu$m is long enough to induce
sizable impact parameter for the charged decay product.
The CMS group has used the impact parameter to study CP property of
the interaction between Higgs and the tau pair~\cite{CMS:2021sdq}.
Here we adopt a similar method proposed in Ref.~\cite{Hagiwara:2016zqz}.
Furthermore, excellent impact parameter resolution can be achieved
by the CEPC vertex detector. The main performance goals for spatial
resolution near the IP can be better than $3~\mu$m~\cite{CEPC-SPPCStudyGroup:2015csa}.
Here, the real impact parameters of the pions are smeared according to Gaussian
distribution with standard deviations $\sigma_{\rm IP} = 3~\mu$m.

We use magnitudes of the $\tau^\pm$ momenta, $|\vec{p}_{\tau^{\pm}}|$,
as the free parameters for finding the best fit.
For single tau decay, for instance $\tau^{-}\to\pi^{-}\nu_{\tau}$,
the opening angle between $\tau^-$ and $\pi^-$ is
\bee
\cos\theta_{\tau^{-}\pi^{-}}
=
\frac{ 2   E_{\tau^{-}}   E_{\pi^{-}} - m_{\tau}^2 - m_{\pi^{-}}^2 }
{ 2 |\vec{p}_{\tau^{-}}| |\vec{p}_{\pi^{-}}| }\,,
\ene
where $E_{\tau^{-}} = \sqrt{ m_\tau^2 + |\vec{p}_{\tau^{-}}|^2 }$ given by the on-shell
condition. Then, the momentum of $\tau^-$ is given as
\bee\label{eq:ptau}
\vec{p}_{\tau^{-}}
=
|\vec{p}_{\tau^{-}}| \cdot
 \frac{ \vec{b}_{\pi^{-}} + \frac{ |\vec{b}_{\pi^{-}}| } { \tan\theta_{\tau^{-}\pi^{-}} } \frac{\vec{p}_{\pi^{-}} }{ |\vec{p}_{\pi^{-}}| } } { \bigg|\vec{b}_{\pi^{-}} + \frac{ |\vec{b}_{\pi^{-}}| } { \tan\theta_{\tau^{-}\pi^{-}} } \frac{\vec{p}_{\pi^{-}} }{ |\vec{p}_{\pi^{-}}| } \bigg| }\,,
\ene
where $\vec{b}_{\pi^{-}}$ is the impact parameter of the $\pi^-$.
Momentum of the neutrino can be obtained by the momentum conservation
condition, $ p_{\nu_\tau}^\mu =  p_{\tau^{-}}^\mu - p_{\pi^{-}}^\mu$.
Similarly, one can obtain momenta of the $\tau^+$ and anti-neutrino as
functions of the parameter $|\vec{p}_{\tau^{+}}|$ and the impact parameter
$\vec{b}_{\pi^{+}}$.
The best values of the parameters $|\vec{p}_{\tau^{\pm}}|$ are obtained by
minimizing following likelihood function
\bee
L
=
L_{BW}(\hat{m}_{\tau\tau}, m_Z, \varGamma_Z) \cdot
L_{G}(\hat{m}_{\tau\tau} - m_Z, \varGamma_Z) \cdot
\prod_{\mu = 0,1,2,3} L_{G}( \hat{p}_{Z}^{\mu} - p_Z^{\mu}, \sigma_Z^{\mu})\,,
\ene
where $L_{BW}$ is the usual Breit-Wigner distribution for the resonant
production of the $Z$ boson, and $\hat{m}_{\tau\tau}$ is the reconstructed
invariant mass of the tau-lepton pair; $\hat{p}_{Z}^{\mu}$ is the reconstructed
momentum of the $Z$ boson, and $p_Z^{\mu}$ is the momentum obtained by
summing momenta of its decay product; $L_{G}(x, y)$ is the Gaussian function
with mean value $x$ and variance $y$. Here $\sigma_Z^{\mu}$ is estimated by
our numerical simulation.

\begin{figure}[htb!]
\centering
\includegraphics[width=0.32\textwidth]{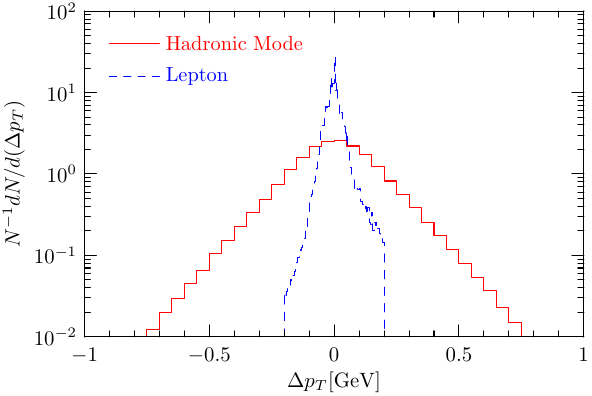}
\includegraphics[width=0.32\textwidth]{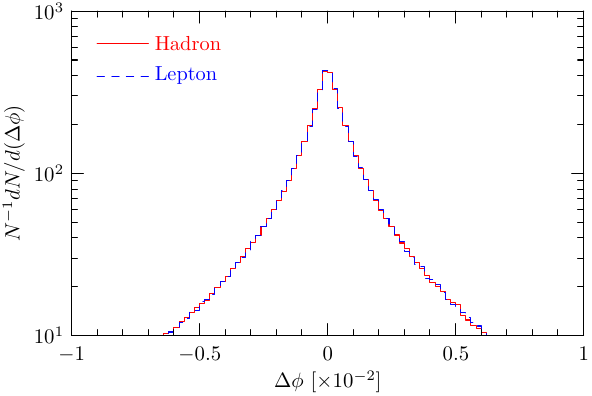}
\includegraphics[width=0.32\textwidth]{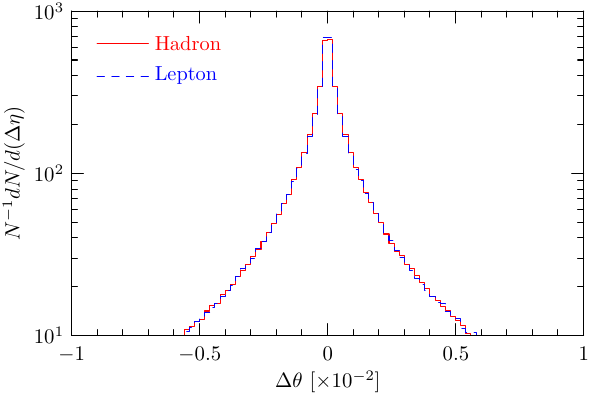}
\caption{
Normalized number of events as a function of $\Delta p_T$ (left), $\Delta \phi$ (middle) and $\Delta \eta$ (right) for the reconstructed $\tau$ lepton in leptonic (blue) and hadronic (red) decay modes of $Z$ boson.
}
\label{fig:recon1}
\end{figure}

\begin{figure}[htb!]
\centering
\includegraphics[width=0.52\textwidth]{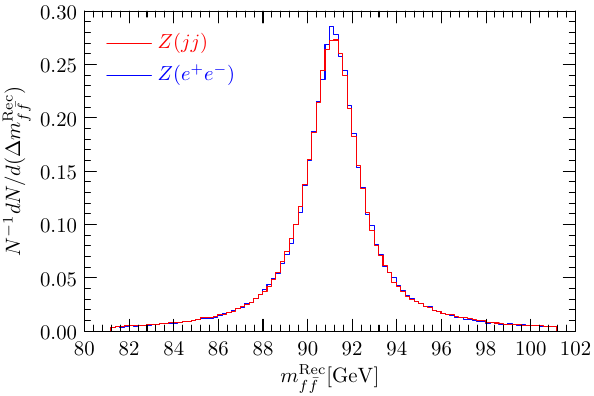}
\caption{
Normalized number of events as a function of reconstructed $Z$ mass $m_{f\bar{f}}^{\rm Rec}$ in leptonic (red) and hadronic (blue) decay modes of $Z$ boson.
}
\label{fig:recon2}
\end{figure}

\begin{figure}[htb!]
\centering
\includegraphics[width=0.36\textwidth]{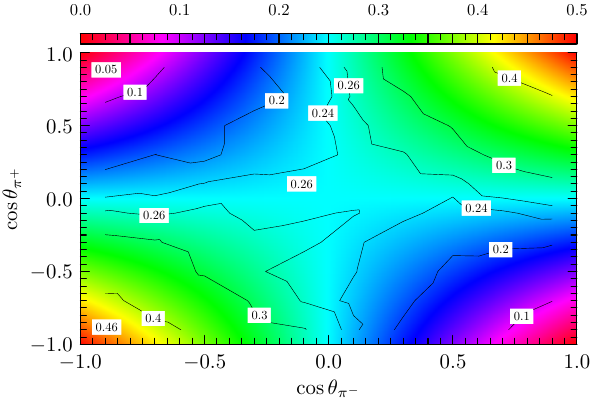}
\includegraphics[width=0.36\textwidth]{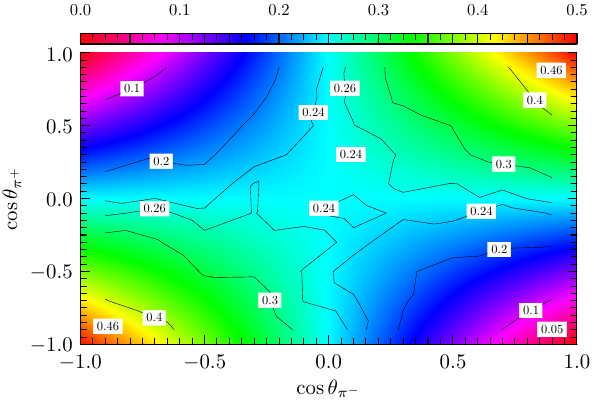}\\[2mm]
\includegraphics[width=0.36\textwidth]{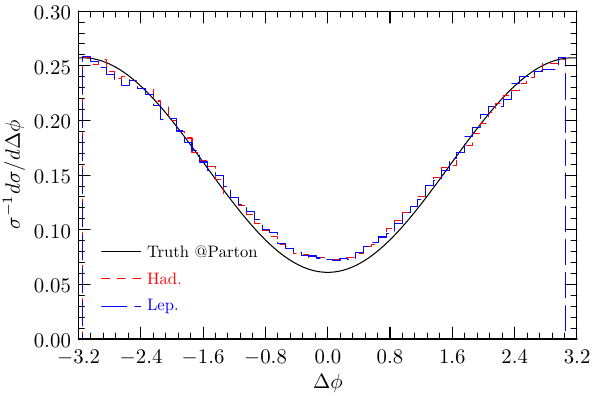}
\caption{
Reconstructed longitudinal (top) and transverse (bottom) correlations for hadronic (top left) and leptonic (top right) decay modes of $Z$ boson.
}
\label{fig:recon}
\end{figure}

In Figs.~\ref{fig:recon1} and \ref{fig:recon2}, after using the method of impact parameters, we show the observable uncertainties for the reconstructed $\tau$ lepton and the reconstructed $Z$ mass in different decay modes of $Z$ boson, respectively. It turns out that the reconstruction in hadronic mode is as good as that in leptonic mode for azimuthal angle, rapidity and the reconstructed $Z$ boson mass. The reconstructed transverse momentum $p_T$ in hadronic mode is still worse than that in leptonic mode.
We also display the reconstructed longitudinal and transverse correlations for hadronic and leptonic decay modes of $Z$ in Fig.~\ref{fig:recon}. As seen in the bottom panel, the reconstructed transverse correlation agrees well with the true result at parton level. It turns out that the method of impact parameters has no the drawback of two-fold ambiguity.

\begin{figure}[htb!]
\centering
\includegraphics[width=0.66\textwidth]{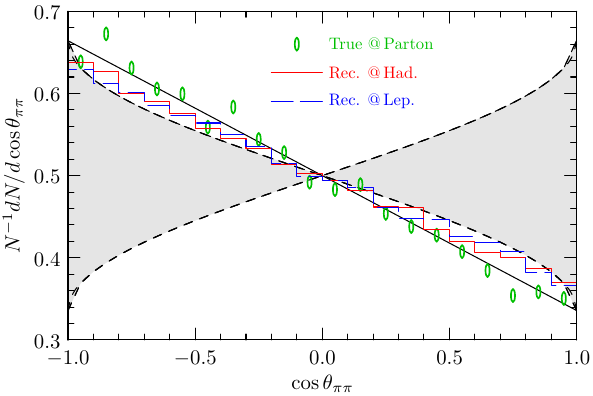}
\caption{Reconstructed distributions of $\cos\theta_{\pi\pi}$ for T\"{o}rnqvist's test of Bell inequality. The gray-fitted region
is the phase space consistent with classical prediction.
}
\label{fig:Tornqvist}
\end{figure}

Finally, in Fig.~\ref{fig:Tornqvist}, we show the distribution of $\cos\theta_{\pi\pi}$ for hadronic (red) and leptonic (blue) decay modes of $Z$ and compare with the Bell inequality. The LHVT holds between the two dashed lines (gray fitted region) given by the inequality in Eq.~(\ref{eq:TtestLHVT}). The black solid line shows the normalized differential cross section in Eq.~(\ref{eq:TtestQM}) and the result at parton level is shown in green dots. The simulation results after using the method of impact parameters are represented by the red and blue histograms for hadronic and leptonic decay modes of $Z$ boson, respectively. Naively seen from the distributions, they stay outside the region satisfying the Bell inequality and agree with the QM/QFT prediction in black solid line.

\begin{figure}[htb!]
\centering
\includegraphics[width=0.32\textwidth]{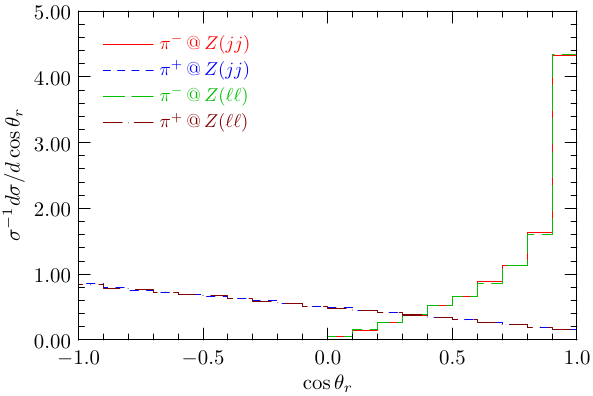}
\includegraphics[width=0.32\textwidth]{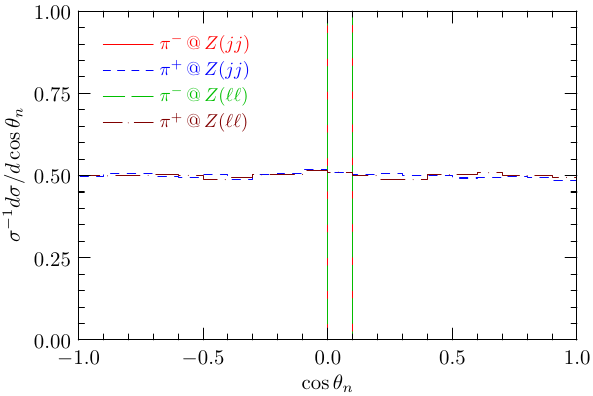}
\includegraphics[width=0.32\textwidth]{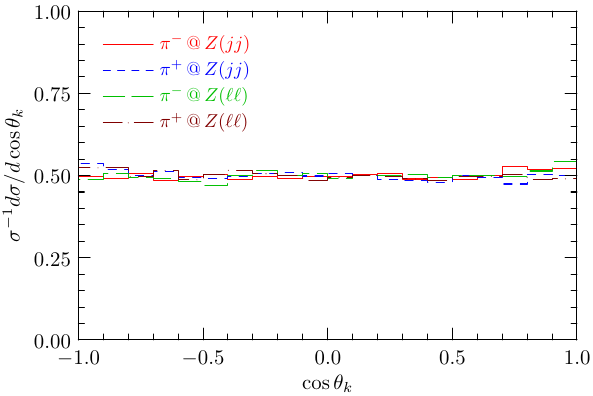}
\caption{
Reconstructed angular distributions of $\pi^\pm$ for $\cos\theta_{\pi,r}$ (left), $\cos\theta_{\pi,n}$ (middle) and $\cos\theta_{\pi,k}$ (right).
}
\label{fig:CHSH}
\end{figure}

Fig.~\ref{fig:CHSH} shows the reconstructed angular distributions of the charged pions.
In general, quantum entanglement disappears when we observe any single angular observable
of the six angles. This is because quantum correlations among $\pi^+$ and $\pi^-$ are integrated out,
as shown by the flat distributions in the middle panel for the $\pi^+$ and the right panel for both
$\pi^+$ and $\pi^-$.
Since the $x-z$ plane is by definition spanned by the momentum direction of the $\tau^-$ and
$\pi^-$, the observable $\cos\theta_{\pi^-,n}$ can only be zero as shown in the middle panel.
The nontrivial distributions shown in the left-panel of the Fig.~\ref{fig:CHSH} are purely kinematic. Since $d\sigma/d\cos\theta_{\pi^-,k}$ is proportional to a constant and $\theta_{\pi^-,r} = {\pi\over 2} - \theta_{\pi^-,k}$, we have
\begin{equation}
  \frac{\sigma^{-1} d\sigma}{ d\cos\theta_{\pi^-,r} }
    =
    \frac{\sigma^{-1} d\sigma}{ d\sin\theta_{\pi^-,k} }
    \propto
    {\rm const.} \times \cot\theta_{\pi^-,r} \,,
\end{equation}
which is essentially reconstructed in our approach as shown by the red-solid and
green-dashed lines in the left-panel of the Fig.~\ref{fig:CHSH}. Similarly,
the asymmetric distribution for the $\pi^+$ is due to the fact that we have defined
$\cos\theta_{\pi-,r} > 0$, which can give a nontrivial integration
on the differential cross section
$d\sigma/d\cos\theta_{\pi-,r}d\cos\theta_{\pi+,r} \propto 1 + C_{rr} \cos\theta_{\pi-,r}\cos\theta_{\pi+,r}$ (the integration region is limited to
the range $\cos\theta_{\pi-,r} \in [0,\, 1]$).

One should note that the calculated coefficients $C_{ij}$ are
not really the spin correlation coefficients of the $\tau$-lepton pair. The polar angles
in Eq.~(\ref{eq:xsec}) are defined for the $\pi$s rather than $\tau$s.
This is a general property of testing Bell-type inequality at colliders,
because the spin (or helicity state) of the particle under consideration can not be
measured directly. The helicity state can only be inferred partially via the angular distribution
of its decay products. For instance, the decay process $\tau\to\pi\nu$ is a good
channel to infer the polarization of the mother $\tau$-lepton.

\section{Sensitivity of CEPC to the Bell Inequality Violation}
\label{sec:Sen}

In this section, we show the sensitivity of CEPC to the Bell inequality violation. As stated in Sec.~\ref{sec:recon}, the analytical reconstruction method suffers from the ambiguous two-fold problem etc. We instead adopt the method of impact parameters for the reconstruction of tau leptons as described in Sec.~\ref{sec:impact}.

At the CEPC with $\sqrt{s}=240$\,GeV, the total cross section of the Higgsstrahlung process
is
\bee
\sigma_{Zh} = 196.2 \fb \,.
\ene
A huge number of the Higgs boson events will be produced with an expected integrated luminosity of $\mathcal{L}=5.6~{\rm ab}^{-1}$~\cite{CEPCStudyGroup:2018ghi}. However, since both the branching
ratios $\calb(h\to\tau\tau) = 6.32\%$ and $\calb(\tau\to\pi\nu_\tau) = 10.82\%$
are small, only hundreds of the events are available to test the Bell inequality.
The following kinematic cuts are used to select well-reconstructed events, and match
to the real detector configuration
\begin{eqnarray}
p_T(\ell/j)> 10\,\gev,~|\eta(\ell/j)|< 3,~
\big| m^{\rm Rec.}_{ff} - m_{Z} \big| < 10\,\gev.
\end{eqnarray}
The efficiency for the above kinematic cuts is 0.645 for the decay mode
$Z\to\ell\ell$, and 0.648 for the hadronic decay mode $Z\to jj$.
Furthermore, as we have mentioned, a universal jet reconstruction efficiency 0.8 will be used in our following estimation.
The number of events can be further reduced
by $\tau$-jet identification which is assumed to be 0.8 for a purity closing to 90\% at the CEPC~\cite{CEPCStudyGroup:2018ghi}.
Table~\ref{tab:noe} gives the expected number of events at the CEPC.

\begin{table}[h]
\renewcommand\arraystretch{1.22}
\begin{center}
\begin{tabular}{ccc}
\hline
\makebox[0.38\textwidth]{CEPC (240 GeV, $5.6~{\rm ab}^{-1}$)} & \makebox[0.1\textwidth]{$Z\to\ell\ell$} & \makebox[0.1\textwidth]{$Z\to jj$}
\\ \hline
No. of Events & 55 & 568
\\ \hline
Kin. Cuts and jet reconstruction & 22 & 151
\\ \hline
$\tau$-identification & 14 & 97
\\
\hline
\end{tabular}
\end{center}
\caption{Number of events used to test the Bell inequality at the CEPC.
}
\label{tab:noe}
\end{table}

The experimental sensitivity for
the T\"{o}rnqvist's approach is studied by
defining the following asymmetric observable
\bee
\cala = \frac{N(\cos\theta_{\pi\pi} < 0 ) - N(\cos\theta_{\pi\pi} > 0 ) }{ N(\cos\theta_{\pi\pi} < 0 ) + N(\cos\theta_{\pi\pi} > 0 ) }\,.
\ene
The analytical prediction of the observable gives an upper bound $\cala=0.119$ in LHVT.
The experimental sensitivity at the CEPC is estimated
by performing 10000 pseudo-experiments and in each experiment we generate as many as possible events to predict the central values and the corresponding uncertainties. Then, we scale the uncertainties to the corresponding luminosities.
We obtain $\cala=0.133\pm 0.269$ for $Z\to \ell\ell$ channel and
$\cala=0.137\pm 0.1$ for $Z\to jj$ channel, respectively,
as listed in Table~\ref{tab:asybell}. Smaller uncertainties can be obtained with $\cala=0.133\pm 0.142$ or
$\cala=0.137\pm 0.053$ for updated luminosity $\mathcal{L}=20~{\rm ab}^{-1}$.
Both central values are above the bound in LHVT.
The channel from the hadronic decay mode of $Z$ boson produces more events and gives more reasonable result with small error.
In the CHSH approach, the LHVT supports the fact that the sum of the two largest eigenvalues of the matrix $U=C^TC$ (denoted by $m_1+m_2$) is not larger than 1. It turns out that both channels lead to $m_1+m_2>1$, as listed in Table~\ref{tab:asybell}.
As we can see, for both the T\"{o}rnqvist's and CHSH approachs,
the Bell inequality can only be tested below $1\sigma$ level
at the CEPC .
It is expected that the sensitivity can be further improved by using
sophisticated jet reconstruction method and enhanced $\tau$-jet identification
efficiency. Note that the results in the fourth and fifth columns of Table~\ref{tab:asybell} are from our detector-level simulation. The simulation results of SM expectation change from parton-level to detector-level. The parton-level and detector-level predictions are different because the experimental uncertainties can diminish the magnitudes of the spin correlation effects as a result of imperfect reconstruction (and also reconstruction efficiency).
In this work, we compare the detector-level QM simulation results with the bound in LHVT directly. We will leave the study of a more reasonable comparison in a future work.

\begin{table}[h]
\renewcommand\arraystretch{1.22}
\begin{center}
\begin{tabular}{ccccc}
\hline
\makebox[0.15\textwidth]{Channels}  & \makebox[0.08\textwidth]{Obs.} & 
\makebox[0.1\textwidth]{Clas.} &
\makebox[0.2\textwidth]{Exp. @\,5.6$\ab^{-1}$} &
\makebox[0.2\textwidth]{Exp. @\,20$\ab^{-1}$}
\\ \hline
\multirow{2}{*}{ $Z\to\ell\ell$} & $\cala$
& $\le 0.119$ & $0.133 \pm 0.269$ & $0.133 \pm 0.142$
\\ 
 & $m_1+m_2$
 & $\le1$ & $1.04 \pm 0.921$ &
$1.04 \pm 0.481$
\\ \hline
\multirow{2}{*}{ $Z\to jj$} & $\cala$
& $\le 0.119$ & $0.137 \pm 0.1$ & $0.137 \pm 0.053$
\\ 
 & $m_1+m_2$
& $\le1$ & $1.05 \pm 0.355$ & $1.05 \pm 0.188$
 \\
\hline
\end{tabular}
\end{center}
\caption{The results of observables testing the Bell inequality in T\"{o}rnqvist's method and the CHSH approach. The experimental predictions
are given for the CEPC with colliding energy $\sqrt{s}=240\,\gev$
and total luminosities $5.6~\ab^{-1}$
and $20~\ab^{-1}$.
}
\label{tab:asybell}
\end{table}

It is worthy to point out that, compared to the T\"{o}rnquivst's method,
the calculation of the sum of two largest eigenvalues
of $C^T C$ as an estimator in the CHSH approach needs the estimation of
the spin projections along all the 3 possible independent directions. Hence, it is relatively more difficult for small data sample.
For the T\"{o}rnquivst's method, the bounds given by the Bell inequality on the LHVT are shown by dashed lines in the distribution of $\cos\theta_{\pi\pi}$ in Fig.~\ref{fig:Tornqvist}. In order to give a quantitative determination of the Bell inequality violation, we define the above asymmetric observable $\mathcal{A}$. In this way, the LHVT has a quantitative upper bound as shown in Table~\ref{tab:asybell}.
Different (quantum) models
lead to different values of $\mathcal{A}$ and $m_1 + m_2$, and are certainly affected by new physics.
As stated in Introduction, it is reasonable to assume that both the $h\to \tau\tau$ decay and tau decay processes are described by the SM. The QM prediction in Table~\ref{tab:asybell} is obtained in the SM. We thus claim that the Bell inequality can be tested under the assumption of SM below $1\sigma$ level at the CEPC.

In summary, the QM prediction in Table~\ref{tab:asybell} is given by the parton-level events. The experimental values are based on our Monte-Carlo simulation. There are two factors affecting the disagreement. One is the experimental uncertainty due to the detector effects which can reduce the quantum correlation such
that the result tends to deviate from the QM prediction. The second factor is the reconstruction method. The correlation can not be reconstructed precisely. Even if the number of events is increased, the deviation would still exist.

\section{Discussions and Conclusions}
\label{sec:Con}

Since spin state can not be directly measured at collider, it is a challenge of the test of quantum entanglement and Bell-nonlocality in high-energy collider physics.
However, testing Bell-nonlocality in high energy scattering process is essentially important because it provides a unique way to address the quantum
entanglement at high energy scale.
We investigate the testability of Bell inequality through $h\to \tau^+\tau^-$,
which is an ideal system to observe the LHVT violation,
at future $e^+e^-$ collider CEPC.
We demonstrated how to use angular distributions of decay products of the
spin-correlated $\tau$-pair to address the Bell-nonlocality.
Future $e^+e^-$ colliders can improve the measurement accuracy of the spin correlation of tau lepton pairs from Higgs boson decay.
Two realistic methods of testing Bell inequality, i.e., T\"{o}rnqvist's method and the CHSH inequality are studied
in terms of the polarization correlation in decay $h\to \tau^+\tau^-\to \pi^+\bar{\nu}_\tau \pi^- \nu_\tau$.

We simulate the production of $e^+e^-\to Zh\to Z\tau^+\tau^-$ as well as the $Z$ boson's leptonic and hadronic decay modes. The detector effects of CEPC including uncertainties for tracks and jets from $Z$ boson are taken into account. We also describe necessary reconstruction approaches to measure quantum entanglement between $\tau^+$ and $\tau^-$.
Finally, we find that for both the T\"{o}rnqvist's and CHSH approachs,
the Bell inequality can be tested at the CEPC below $1\sigma$ level.
Further improvements are expected by employing
sophisticated jet reconstruction method and enhanced $\tau$-jet identification
efficiency.

We also noticed that in Ref.~\cite{Altakach:2022ywa} the authors studied the same topic at ILC and FCC-ee.
The expected total number of events is 385 at
ILC (250\,GeV, $\mathcal{L}=3~{\rm ab}^{-1}$) or 663 at FCC-ee (240\,GeV, $\mathcal{L}=5~{\rm ab}^{-1}$). In contrast, our estimation for CEPC (240\,GeV, $\mathcal{L}=5.6~{\rm ab}^{-1}$)
is 111. Even if the total cross section at ILC or FCC-ee is slightly
larger (240.1\,fb or 240.3 fb) than the one (196.2\,fb) at the CEPC,
it is too small to have more than 2 times or 5 times larger number of events.
Such a difference can only come from the detector effects.
The reconstruction of the mode
$e^+e^-\to Z(\to jj) h(\to \tau_\pi\tau_\pi)$, which has the largest production
rate, is rather difficult and thus influences the detector
effects on the reconstruction. Since the jets (from $Z$ or $h$)
are relatively soft at the CEPC (and ILC-250), the usual jet clustering does
not give proper output. Hence, the reconstruction efficiency is rather
low. Some sophisticated clustering method has to be employed.
In our work, we chose the smearing method for CEPC to solve this problem.

In addition, we also proposed to use the T\"{o}rnqvist’s method to test
the Bell inequality at CEPC. The reason is twofold. Firstly, it is much
simple and straightforward. Secondly, in case of the (very) small number of events,
the estimation of the correlation efficiency $C_{ij}$ via the integration
of the phase space is unstable/unsafe, and hence induces a very large
fluctuation of the central/expected value of the observable.
This can be seen in Table~\ref{tab:asybell}, particularly for the
leptonic decay mode of the $Z$ boson.

\section*{Acknowledgements}

T.L. would like to thank Xue-Qian Li for helpful discussion.
T.L. is supported by the National Natural Science Foundation of China (Grants No. 12375096, 12035008, 11975129) and ``the Fundamental Research Funds for the Central Universities'', Nankai University (Grant No. 63196013).
K. M. was supported by the Natural Science Basic Research Program of Shaanxi (Program No. 2023-JC-YB-041), and the Innovation Capability Support Program of Shaanxi (Program No. 2021KJXX-47).

\appendix
\section{The spin correlation coefficients in T\"{o}rnqvist’s method}
\label{Appendix}

For the T\"{o}rnqvist’s method, we calculate the correlation coefficients in the following way.
Assuming the $\tau$ and $\bar\tau$ are polarized along
$\vec{a}$ and $\vec{b}$, respectively, their helicity basis can be chosen to be the eigenstates of the helicity opeators $\vec{a}\cdot\vec\sigma_{\tau}$ and
$\vec{b}\cdot\vec\sigma_{\bar\tau}$, respectively. Then, the expected value
of the correlation matrix
$\mathcal{O}(\vec{a}, \vec{b}) = \big(\vec{a}\cdot\vec\sigma_{\tau}\big)
\big(\vec{b}\cdot\vec\sigma_{\bar\tau}\big)$
is given as
\begin{equation}
    \langle h_\tau h_{\bar\tau}| \rho_{\tau\bar\tau} \, \mathcal{O}(\vec{a}, \vec{b})  | h_\tau h_{\bar\tau}\rangle
    =
    \langle h_\tau h_{\bar\tau}| \frac{1}{4}\big(1 - \vec\sigma_{\tau}\cdot\vec\sigma_{\bar\tau}\big) \,  \big(\vec{a}\cdot\vec\sigma_{\tau}\big)
\big(\vec{b}\cdot\vec\sigma_{\bar\tau}\big)  | h_\tau h_{\bar\tau}\rangle \,,
\end{equation}
where $h_{\tau(\bar\tau)}$ denotes the helicity of $\tau (\bar\tau)$.
By using the following relation
\begin{equation}
    \vec\sigma_{\tau}\cdot\vec\sigma_{\bar\tau}
    =\frac{1}{2}\left( \vec\Sigma^2 - 6 \right) \,,
\end{equation}
where $\vec\Sigma = \vec\sigma_{\tau} + \vec\sigma_{\bar\tau}$ and $\vec\Sigma^2 = 0$ for
spin singlet, we have
\begin{equation}
    \langle h_\tau h_{\bar\tau}| \rho_{\tau\bar\tau} \, \mathcal{O}(\vec{a}, \vec{b})  | h_\tau h_{\bar\tau}\rangle
    =
    \langle h_\tau h_{\bar\tau}| \big(\vec{a}\cdot\vec\sigma_{\tau}\big)
\big(\vec{b}\cdot\vec\sigma_{\bar\tau}\big)  | h_\tau h_{\bar\tau}\rangle
= h_\tau h_{\bar\tau}\,.
\end{equation}
For $\vec{a} = \hat{\vec{e}}_z$ and $ \vec{b} = \hat{\vec{e}}_z$, the spin singlet requires the helicities of $\tau$ and $\bar{\tau}$ satisfy
\begin{equation}
    h=h_\tau + h_{\bar\tau} =0\,.
\end{equation}
Hence, we obtain the correlation coefficients as
\begin{equation}
    C_{zz} = \langle h_\tau h_{\bar\tau}| \rho_{\tau\bar\tau} \, \mathcal{O}(\hat{\vec{e}}_z, -\hat{\vec{e}}_z)  | h_\tau h_{\bar\tau}\rangle
    = h_\tau h_{\bar\tau} = -1\,,
\end{equation}
and similarly
\begin{equation}
    \begin{aligned}
        C_{xx} = \langle h_\tau h_{\bar\tau}| \rho_{\tau\bar\tau} \, \mathcal{O}(\hat{\vec{e}}_x, -\hat{\vec{e}}_x)  | h_\tau h_{\bar\tau}\rangle
    & = -1\,,
    \\[3mm]
    C_{yy} = \langle h_\tau h_{\bar\tau}| \rho_{\tau\bar\tau} \, \mathcal{O}(\hat{\vec{e}}_y, -\hat{\vec{e}}_y)  | h_\tau h_{\bar\tau}\rangle
    & = -1\,.
    \end{aligned}
\end{equation}
The above calculations show that the correlation coefficients in our helicity basis
are $(-1,\, -1,\, -1)$. In the above calculations, the condition $\vec\Sigma^2 = 0$,
which means the $\tau\bar\tau$ system is a spin-singlet, is crucial. One can have different results if the spin projection axis is
defined in a different way.

\bibliography{Belltau}

\end{document}